# Retarded Interaction of Electromagnetic field and Symmetry Violation of Time Reversal in Non-linear Optics

Mei Xiaochun

(Institute of Theoretical Physics in Fuzhou, China, E-mial:mxc001@163.com)

**Abstract** Based on Document (1), by considering the retarded interaction of radiation fields, the third order transition probabilities of stimulated radiations and absorptions of light are calculated. The revised formulas of nonlinear polarizations are provided. The results show that that the general processes of non-linear optics violate time reversal symmetry. The phenomena of non-linear optics violating time reversal symmetry just as sum frequency, double frequency, different frequencies, double stable states, self-focusing and self-defocusing, echo phenomena, as well as optical self-transparence and self absorptions and so on are analyzed.

## 1. Time reversal of the third order process of double photons

As proved in Document (1), after the retarded effect of radiation field is considered, the processes of light's stimulated radiations and absorptions violate time reversal symmetry. In this paper, we continue to discuss the problems of time reversal in the processes of non-linear optics. The problem of double photons is discussed at first. According to the formula (33) in document (1), by considering rotation wave approximation, when $2\omega = \omega_{ml}$ we have the transition probability amplitude of the second order process of double photon stimulated absorption

$$a_m^{(2)}(t)_{2\omega=\omega_{ml}} = \left[\frac{e^{-i(2\omega-\omega_{ml})t}-1}{\hbar(2\omega-\omega_{ml})}\right]\left[\hat{F}_{2ml}^+ + \frac{2\hat{F}_{1ml}^+\left(\hat{F}_{1ll}^+ - \hat{F}_{1mm}^+\right)}{\hbar\omega_{ml}}\right] \qquad (1)$$

So the total transition probability of stimulated absorption of double photons in the first and second processes in unit time is

$$W_{2\omega=\omega_{ml}}^{(2)} = \frac{2\pi}{\hbar^2}\left\{\left|\hat{F}_{2ml}^+\right|^2 + \frac{4}{\hbar^2\omega_{ml}^2}\left|\hat{F}_{1ml}^+\right|^2\left|\hat{F}_{1ll}^+ - \hat{F}_{1mm}^+\right|^2\right.$$

$$\left. + \frac{4}{\hbar\omega_{ml}}Re\left[\left(\hat{F}_{2ml}^+\right)^*\hat{F}_{1ml}^+\left(\hat{F}_{1ll}^+ - \hat{F}_{1mm}^+\right)\right]\right\}\delta(2\omega-\omega_{ml}) \qquad (2)$$

Here "Re" represents the real part of the function. Similarly, we can obtain the time reversal of transition probability of stimulated absorption of double photons in the second process

$$a_{Tm}^{(2)}(t)_{2\omega=\omega_{ml}} = -\frac{\left[e^{-i(2\omega-\omega_{ml})t}-1\right]}{\hbar(2\omega-\omega_{ml})}\left[\hat{F}'^{+}_{2ml} + \frac{2\hat{F}'^{+}_{1ml}\left(\hat{F}'^{+}_{1ll}-\hat{F}'^{+}_{1mm}\right)}{\hbar\omega_{ml}}\right] \quad (3)$$

By relations $\left|\hat{F}'^{+}_{2ml}\right|^2 = \left|\hat{F}_{2ml}\right|^2$, $\left|\hat{F}'^{+}_{1ml}\right|^2 = \left|\hat{F}_{1ml}\right|^2$, $\hat{F}'_{1ml} = \left(\hat{F}^{+}_{1ml}\right)^*$, $\hat{F}^{+}_{1ml} = \hat{F}^{*}_{1ml}$, $\hat{F}'_{2ml} = \left(\hat{F}^{+}_{2ml}\right)^*$ and $\hat{F}'^{+}_{2ml} = \hat{F}^{*}_{2ml}$, the time reversal of total transition probability of stimulated absorption of double photons in the first and second processes is

$$W_{T2\omega=\omega_{ml}}^{(2)} = \frac{2\pi}{\hbar^2}\left\{\left|\hat{F}_{2ml}\right|^2 + \frac{4}{\hbar^2\omega_{ml}^2}\left|\hat{F}_{1ml}\right|^2\left|\hat{F}_{1ll}-\hat{F}_{1mm}\right|^2\right.$$

$$\left. + \frac{4}{\hbar\omega_{ml}}\text{Re}\left[\hat{F}^{*}_{2ml}\hat{F}_{1ml}\left(\hat{F}_{1ll}-\hat{F}_{1mm}\right)\right]\right\}\delta(2\omega-\omega_{ml}) \quad (4)$$

According to the formulas (18), (19) and (32) in Document (1), we have

$$\left(\hat{F}^{+}_{2ml}\right)^*\hat{F}^{+}_{1ml}\left(\hat{F}^{+}_{1ll}-\hat{F}^{+}_{1mm}\right) = \hat{F}^{*}_{2ml}\hat{F}_{1ml}\left(\hat{F}_{1ll}-\hat{F}_{1mm}\right) \quad (5)$$

as well as $\left|\hat{F}_{2ml}\right|^2 = \left|\hat{F}^{+}_{2ml}\right|^2$, $\left|\hat{F}_{1ml}\right|^2 = \left|\hat{F}^{+}_{1ml}\right|^2$, $\left|\hat{F}^{+}_{1ll}-\hat{F}^{+}_{1mm}\right|^2 = \left|\hat{F}_{1ll}-\hat{F}_{1mm}\right|^2$, we have $W_{T2\omega=\omega_{ml}}^{(2)} = W_{2\omega=\omega_{ml}}^{(2)}$.

So there is no symmetry violation of time reversal in the first and second order processes of double photons. We should consider the third order process. In light of (12) in Document (1), the transition probability amplitude of the third order process is

$$a_m^{(3)}(t) = \frac{1}{i\hbar}\sum_n\int_0^t \hat{H}'_{2mn}a_n^{(1)}(t)e^{i\omega_{mn}t}dt + \frac{1}{i\hbar}\sum_n\int_0^t \hat{H}'_{1mn}a_n^{(2)}(t)e^{i\omega_{mn}t}dt \quad (6)$$

By taking the integral of the formula, we can obtain the probability amplitude of the third order processes. The result is shown in appendix. For the double photon absorption processes with $2\omega = \omega_{ml}$, by rotation wave approximation, the probability amplitude and transition probability in unit time are individually

$$a_m^{(3)}(t)_{2\omega=\omega_{ml}} = \frac{\left[e^{-i(2\omega-\omega_{ml})t}-1\right]}{\hbar(2\omega-\omega_{ml})}\frac{4\hat{F}^{+}_{1ml}\left(\hat{F}^{+}_{1ll}-\hat{F}^{+}_{1mm}\right)\left(\hat{F}^{+}_{1ll}-\hat{F}^{+}_{1ll}\right)}{\hbar^2\omega_{ml}^2} \quad (7)$$

$$W_{2\omega=\omega_{ml}}^{(3)} = \frac{2\pi}{\hbar^2}\left\{\left|\hat{F}^{+}_{2ml}\right|^2 + \frac{4(1+4A_l^2)}{\hbar^2\omega_{ml}^2}\left|\hat{F}^{+}_{1ml}\right|^2\left|\hat{F}^{+}_{1ll}-\hat{F}^{+}_{1mm}\right|^2\right.$$

$$\left. + \frac{4}{\hbar\omega_{ml}}\text{Re}\left[(1+i2A_l)\left(\hat{F}^{+}_{2ml}\right)^*\hat{F}^{+}_{1ml}\left(\hat{F}^{+}_{1ll}-\hat{F}^{+}_{1mm}\right)\right]\right\}\delta(2\omega-\omega_{ml}) \quad (8)$$

Here $iA_l = \hat{F}_{1ll} - \hat{F}^{+}_{1ll}$. On the other hand, by taking $k \to j$ in (24) in Document (1), we can get the time reversal of (6)

$$a_{Tm}^{(3)}(t) = a_l^{(3)}(-t) = -\frac{1}{i\hbar}\sum_j \int_0^t \hat{H}'_{2Tlj} a_j^{(1)}(-t) e^{i\omega_{jl}t} dt - \frac{1}{i\hbar}\sum_j \int_0^t \hat{H}'_{1Tlj} a_j^{(2)}(-t) e^{i\omega_{jl}t} \qquad (9)$$

By considering relations $\hat{H}'_{2Tlj} = \hat{H}'^{*}_{2jl}$, $\hat{H}'_{1Tlj} = \hat{H}'^{*}_{1jl}$, $\hat{F}'_{1ml} = (\hat{F}^+_{1ml})^*$, $\hat{F}'^+_{1ml} = \hat{F}^*_{1ml}$, $\hat{F}'_{2ml} = (\hat{F}^+_{2ml})^*$ and $\hat{F}'^+_{2ml} = \hat{F}^*_{2ml}$ as well as by the same method to do integral and take rotation wave approximation,, we get the time reversals of probability amplitude and transition probability in the third order process of double photon absorption individually

$$a_{Tm}^{(3)}(t)_{2\omega=\omega_{ml}} = \frac{\left[e^{-i(2\omega-\omega_{ml})t} - 1\right]}{\hbar(2\omega-\omega_{ml})} \frac{4\hat{F}'^+_{1ml}(\hat{F}'^+_{1ll} - \hat{F}'^+_{1mm})(\hat{F}'_{1mm} - \hat{F}'^+_{1mm})}{\hbar^2 \omega_{ml}^2}$$

$$= -\frac{\left[e^{-i(2\omega-\omega_{ml})t} - 1\right]}{\hbar(2\omega-\omega_{ml})} \frac{4\hat{F}^*_{1ml}(\hat{F}^*_{1ll} - \hat{F}^*_{1mm})(\hat{F}_{1mm} - \hat{F}^+_{1mm})^*}{\hbar^2 \omega_{ml}^2} \qquad (10)$$

$$W^{(3)}_{T2\omega=\omega_{ml}} = \frac{2\pi}{\hbar^2}\left\{\left|\hat{F}_{2ml}\right|^2 + \frac{4(1+4A_m^2)}{\hbar^2\omega_{ml}^2}\left|\hat{F}_{1ml}\right|^2\left|\hat{F}_{1ll} - \hat{F}_{1mm}\right|^2 \right.$$

$$\left. + \frac{4}{\hbar\omega_{ml}} Re\left[(1+i2A_m)\hat{F}_{2ml}\hat{F}^*_{1ml}(\hat{F}^*_{1ll} - \hat{F}^*_{1mm})\right]\right\}\delta(2\omega - \omega_{ml}) \qquad (11)$$

Here $iA_m = \hat{F}_{1mm} - \hat{F}^+_{1mm}$. Comparing with (8), we know that the difference $A_l \neq A_m$ leads to time reversal symmetry violation. But if the high order multiple moment effects are omitted, symmetry violations of time reversals in the third order processes would not exist.

## 2. Sum frequency process and its time reversal

The process of sum frequency process in non-linear optics is that an electron translates into higher energy lever $|m\rangle$ from low energy level $|l\rangle$ by absorbing two photons with frequencies $\omega_1$ and $\omega_2$ individually, then emits out a photon with frequency $\omega_3 = \omega_1 + \omega_2$ and translates from higher energy level $|m\rangle$ into low energy level $|l\rangle$ again. Suppose that incident light is parallel one containing frequencies $\omega_1$, $\omega_2$ and $\omega_3 = \omega_1 + \omega_2$, and the strength of electrical field is $\vec{E}_0$, the interaction Hamiltonians between electron and radiation field are

$$\hat{H}'_1 = \sum_{\lambda=1}^3 \left(\hat{F}_{1\lambda}e^{i\omega_\lambda t} + \hat{F}^+_{1\lambda}e^{-i\omega_\lambda t}\right) \qquad \hat{H}'_2 = \sum_{\lambda=1}^3 \left(\hat{F}_{2\lambda}e^{i\omega_\lambda t} + \hat{F}^+_{1\lambda}e^{-i\omega_\lambda t} + \hat{F}_{0\lambda}\right) \qquad (12)$$

Here

$$\hat{F}_{1\lambda} = -\frac{q\vec{E}_0}{2\omega_\lambda \mu} \cdot e^{-i\vec{k}_\lambda \cdot \vec{R}}\hat{p} \qquad \hat{F}^+_{1\lambda} = -\frac{q\vec{E}_0}{2\omega_\lambda \mu} \cdot \hat{p}^+ e^{i\vec{k}_\lambda \cdot \vec{R}} \qquad (13)$$

$$\hat{F}_{2\lambda} = \frac{q^2 E_0^2}{2\omega_\lambda^2 \mu} e^{-i2\vec{k}_\lambda \cdot \vec{R}} \qquad \hat{F}_{2\lambda}^+ = \frac{q^2 E_0^2}{2\omega_\lambda^2 \mu} e^{i2\vec{k}_\lambda \cdot \vec{R}} \qquad \hat{F}_{0\lambda} = \frac{q^2 E_0^2}{\omega_\lambda^2 \mu} \qquad (14)$$

When we calculate probability amplitude, the result corresponds to let $\omega \to \omega_\lambda$ and take sum over index $\lambda$ in (33) in Document (1) or in the formula in the appendix of this paper. For the process an electron absorbs two photons with frequencies $\omega_1$ and $\omega_2$, transits from low energy level $|l\rangle$ into high energy level $|m\rangle$, transition probability corresponds to let $2\omega = \omega_1 + \omega_2 = \omega_{ml}$ in (8) of double photon absorption process. When the electron transits back from high energy level $|m\rangle$ into low energy level $|l\rangle$ by emitting a photon with frequency $\omega_3$, after the retarded effect and high order processes are considered, according to (47) in Document (1), the transition probability is

$$W^{(2)}_{\omega_3 = \omega_{ml}} = \frac{2\pi}{\hbar^2} |\hat{F}_{1ml}|^2 \left\{ 1 + \frac{A_m'^2}{\hbar^2 \omega_{ml}^2} \right\} \delta(\omega_3 - \omega_{ml}) \qquad (15)$$

Therefore, for the sum frequency process that an electron translates into state $|m\rangle$ from state $|l\rangle$ by absorbing two photons with frequencies $\omega_1$ and $\omega_2$, then translates back into original state $|l\rangle$ from state $|m\rangle$ by emitting out a photon with frequencies $\omega_3 = \omega_1 + \omega_2$, the total transition probability is

$$W^{(2)}_{\omega_1 + \omega_2 = \omega_{ml}} + W^{(2)}_{\omega_3 = \omega_{ml}} = \frac{2\pi}{\hbar^2} \left\{ |\hat{F}_{2ml}^+|^2 + \frac{4(1 + 4A_l^2)}{\hbar^2 \omega_{ml}^2} |\hat{F}_{1ml}^+|^2 |\hat{F}_{1ll}^+ - \hat{F}_{1mm}^+|^2 \right.$$

$$\left. + \frac{4}{\hbar \omega_{ml}} Re\left[ (1 + i2A_l)(\hat{F}_{2ml}^+)^* \hat{F}_{1ml}^+ (\hat{F}_{1ll}^+ - \hat{F}_{1mm}^+) \right] \right\} \delta(2\omega - \omega_{ml})$$

$$+ \frac{2\pi}{\hbar^2} |\hat{F}_{1ml}|^2 \left\{ 1 + \frac{A_m'^2}{\hbar^2 \omega_{ml}^2} \right\} \delta(\omega_3 - \omega_{ml}) \qquad (16)$$

Meanwhile, according the calculation in (38) in Document (1), after the retarded effects and high order processes are considered, the transition probability that an electron transits from state $|l\rangle$ into state $|m\rangle$ by absorbing a photon with frequency $\omega_3$ is

$$W^{(2)}_{T\omega_3 = \omega_{ml}} = \frac{2\pi}{\hbar^2} |\hat{F}_{1ml}|^2 \left\{ 1 + \frac{A_l^2}{\hbar^2 \omega_{ml}^2} \right\} \delta(\omega_3 - \omega_{ml}) \qquad (17)$$

So for the time reversal of sum frequency process that an electron absorbs a photon with frequency $\omega_3$ and transits from state $|l\rangle$ into state $|m\rangle$, then emits two photons with frequencies $\omega_1$ and $\omega_2$, and transits back into original state $|l\rangle$, the total transition probability is

$$W^{(2)}_{T\omega_1 + \omega_2 = \omega_{ml}} + W^{(2)}_{T\omega_3 = \omega_{ml}} = \frac{2\pi}{\hbar^2} \left\{ |\hat{F}_{2ml}|^2 + \frac{4(1 + 4A_m'^2)}{\hbar^2 \omega_{ml}^2} |\hat{F}_{1ml}|^2 |\hat{F}_{1ll} - \hat{F}_{1mm}|^2 \right.$$

$$\left. + \frac{4}{\hbar \omega_{ml}} Re\left[ (1 + i2A_m') \hat{F}_{2ml} \hat{F}_{1ml}^* (\hat{F}_{1ll}^* - \hat{F}_{1mm}^*) \right] \right\} \delta(2\omega - \omega_{ml})$$

$$+\frac{2\pi}{\hbar^2}\left|\hat{F}_{1ml}^+\right|^2\left\{1+\frac{A_l^2}{\hbar^2\omega_{ml}^2}\right\}\delta(\omega_3-\omega_{ml}) \tag{18}$$

Similarly, because of $A_l \neq A'_m$, sum frequency process violates time reversal symmetry.

By the same method, we can prove that the other processes of non-linear optics just as double frequency, difference frequency, parametric amplification, Stimulated Raman scattering, Stimulated Brillouin scattering and so on are also asymmetric under time reversal. The reason is the same that the light's high order stimulated radiation and absorption processes are asymmetric under time reversal after retarded effect of radiation fields are taken into account.

What is discussed above is based on quantum mechanics. But in the most practical problems of non-linear optics, we calculate the problems based on classical equations of electromagnetic fields. So we need to discuss the revised non-linear polarizations when the retarded effect of radiation field is taken into account. According to the current theory of nonlinear optics, polarizations are regarded unchanged under time reversal. This does not coincident with real observations. The reason is that we only consider the dipolar approximation without considering the retarded effects of radiation field and the high order processes of light's stimulated radiation and absorption. We now discuss the revision of non-linear polarizations after the retarded effect of radiation fields and the high order perturbation processes are taken into account, as well as a great number of phenomena violating symmetry of time reversal in non-linear optics.

## 3. Non-linear polarizations and its symmetry violation of time reversal

In light of (54) in Document (54), we let $\vec{D}'_{ml} = \sqrt{1+\lambda_{ml}}\vec{D}_{ml}$ represent the revised dipolar moment after retarded effect is considered, $\vec{D}'_{lm} = \sqrt{1+\lambda_{lm}}\vec{D}_{lm}$ represent the time reversal of $\vec{D}'_{ml}$. We have $\lambda_{ml} \neq \lambda_{lm}$ and $\vec{D}'_{ml} \neq \vec{D}'_{lm}$ in general. Therefore, as long as we let $\vec{D}_{ml} \to \vec{D}'_{ml}$ in the current formula $\chi^{(n)}_{ij...k}$ of non-linear polarizations, we obtain the revised formula after the retarded effect of radiation fields and the high order perturbation processes are taken into account. Correspondingly, we let $\vec{D}_{lm} \to \vec{D}'_{lm}$ and obtain the time reversal formula $\chi^{(n)}_{Tij...k}$. It is obvious that non-linear polarizations can not keep unchanged under time reversal. For example, for the non-linear polarizations of the second order processes, we have

$$\chi^{(2)}_{ijk}=(-\omega_1-\omega_2,\omega_1,\omega_2)=\frac{N}{4\varepsilon_0\hbar^2}\sum_{nn'}\sum_p\frac{(D'_i)_{\beta n'}(D'_j)_{n'n}(D'_k)_{n'\beta}}{(\omega_{n'\beta}+\omega_2)(\omega_{n\beta}+\omega_1+\omega_2)} \tag{19}$$

Its time reversal is

$$\chi^{(2)}_{Tijk}=(-\omega_1-\omega_2,\omega_1,\omega_2)=\frac{N}{4\varepsilon_0\hbar^2}\sum_{nn'}\sum_p\frac{(D'_i)_{n'\beta}(D'_j)_{n\,n'}(D'_k)_{\beta\,n'}}{(\omega_{n'\beta}+\omega_2)(\omega_{n\beta}+\omega_1+\omega_2)} \tag{20}$$

In general, we have $\vec{D}'_{ml} \neq \vec{D}'_{lm}$ and $\chi^{(2)}_{ijk} \neq \chi^{(2)}_{Tijk}$. So in general situations we have $\chi^{(n)}_{ij...k} \neq \chi^{(n)}_{Tij...k}$. Therefore, the polarization formula of electrical medium and its time reversal are

$$\vec{P}=\varepsilon_0\left(\chi^{(1)}_i E_i + \chi^{(2)}_{ij}E_i E_j + \chi^{(3)}_{ijk}E_i E_j E_k \cdots\right) \tag{21}$$

$$\vec{P}_T = \varepsilon_0 \left( \chi^{(1)}_{Ti} E_i + \chi^{(2)}_{Tij} E_i E_j + \chi^{(3)}_{Tijk} E_i E_j E_k \cdots \right) \tag{22}$$

We have $P_T \neq P$ in general. So the motion equation of classical electrical field and its time reversal are also asymmetrical in general with forms

$$\nabla^2 \vec{E} - \mu_0 \varepsilon_0 \frac{\partial^2}{\partial t^2} \vec{E} = \mu_0 \frac{\partial^2}{\partial t^2} \vec{P} \qquad \nabla^2 \vec{E} - \mu_0 \varepsilon_0 \frac{\partial^2}{\partial t^2} \vec{E} = \mu_0 \frac{\partial^2}{\partial t^2} \vec{P}_T \tag{23}$$

Because (21) and (23) are the basic equations of nonlinear optics, we can say that the general nonlinear optical processes violate time reversal symmetry. In fact, only by analyzing nonlinear optics phenomena without complex calculations, we can show the irreversibility of nonlinear optical processes directly under time reversal. Thought the irreversibility concept of processes is not completely the same with that of the asymmetry of time reversal, they are coincident in essence. So let's analyze same practical examples to exposure further the irreversibility of nonlinear optics processes below.

## 4. Irreversibility of nonlinear optics processes

As we know that the processes of linear optics just as light's propagations, reflection, refraction, polarization and so on in uniform mediums are reversible. For example, light's focusing caused by a common convex mirror at point $A$ shown in Fig.5. When a beam of common light is projected into a convex mirror, it would be focused at focus $O$. If we put a same convex mirror at point $B$, and $O$ is also the focus of convex mirror $B$, the light sent out from $O$ point would become a beam of parallel light again when it transit out convex mirror $B$. The process that light moving from $O \to B$ can be regarded as the time reversal process of light moving from $A \to O$. It is obvious that the process is reversible. The second example is that a beam of white sunlight can be resolved into a spectrum with different colors by a prism. When these lights with difference colors are reflected back into prism along same paths, white sunlight would be formed again. The third example is that a beam of light can become two different polarization lights with different propagation directions when the light is projected into a double refraction crystal. If these two polarization lights are reflected back into the crystal along same path again, the original light is formed. All of these processes are reversible. But in the processes of non-linear optics, reversibility does not exist. Some examples are shown below.

### 4.1 Light's multiple frequency, difference frequency and parameter amplification

A beam of laser with frequency $\omega$ is projected into a proper medium and proper phase matching technology is adopted. The light with multiple frequencies $2\omega$ would be found in penetrating light besides original light with frequency $\omega$ as shown in Fig.1. If the lights with frequencies $\omega$ and $2\omega$ are reflected back into the same medium, as shown in Fig.2, they can't be completely synthesized into the original light with a single frequency $\omega$. Some light with frequency $\omega$ would become multiple frequency light again by multiple frequency process. Some light with frequency $2\omega$ would become the light with frequency $\omega$ by difference frequency process. Meanwhile, some light with frequencies $\omega$ and $2\omega$ would penetrate medium without being changed as shown in Fig 2. So the original input light can' be recovered and the reversibility of process is broken. The situations are the same for the sum frequency, difference frequency and parameter amplification processes and so on.

## 4.2 Bistability of optics [2]

As shown in Fig.3 and Fig.4, the processes of optical bistability are just as the polarization and magnetization processes of ferroelectrics and ferromagnetic. In the processes the hysteretic loops are formed between incident and outgoing electrical field strengths. In the polarization and magnetization processes of ferroelectrics and ferromagnetics, electromagnetic fields changing along positive directions can be regarded as the time reversal of fields changing along negative directions. There exist electric and phase hysteresis. The hysteretic loops are similar to heat engine cycling loops. After a cycling, heat dissipation is produced and the reversibility of process is violated.

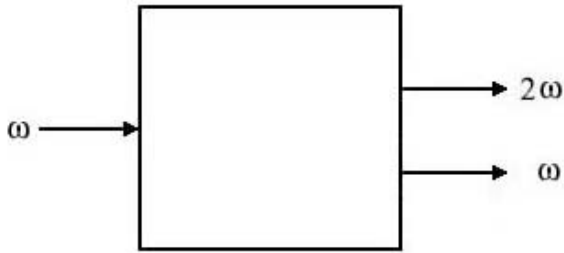 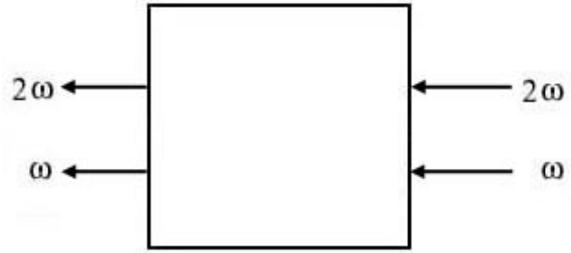

Fig. 1. Process of light's multiple frequency        Fig. 2. Time Reversal process of
                                                      Light's multiple frequency

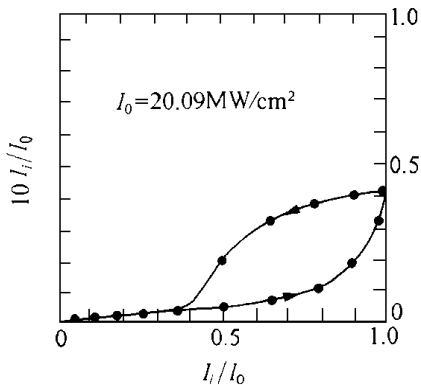 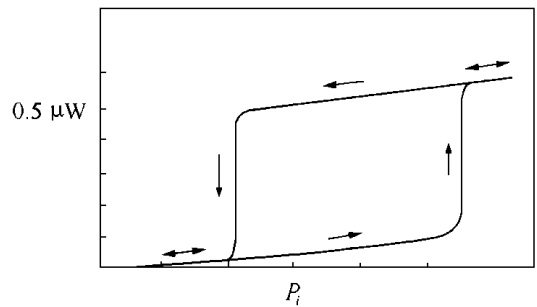

Fig 3. Optical bistability of nitrobenzene        Fig 4. Optical bistability of mixing type

## 4.3 Self-focusing and self-defocusing processes of light [3]

Medium's refractive index would be changed nonlinearly when a beam of laser with uneven distribution on its cross section just as Gauss distribution is projected into a proper medium. The result is that medium seems to become a convex or concave mirror so that parallel light would be focused or defocused. This is just self-focusing and self-defocusing of light. The stationary self-focusing process is shown in Fig.6. Parallel light is focused at point $O$. Then it becomes a thin beam of light projecting out medium. Comparing it with common focusing process shown in Fig.5, if the self-focusing process is reversible, the light focused at pint $O$ would become parallel light again when it projecting out the medium as shown in dotted lines in Fig 6. But it dose not actually. So self-focusing processes are irreversible, and so do for self-defocusing process of light.

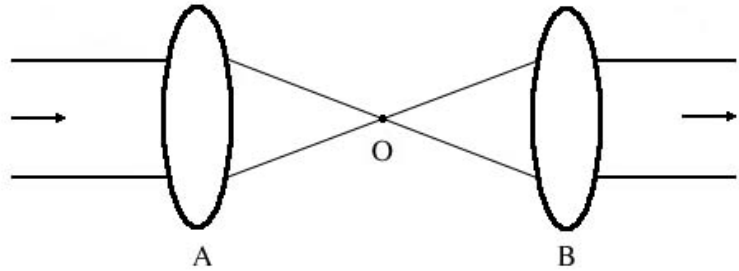

Fig 5. Focusing process of light

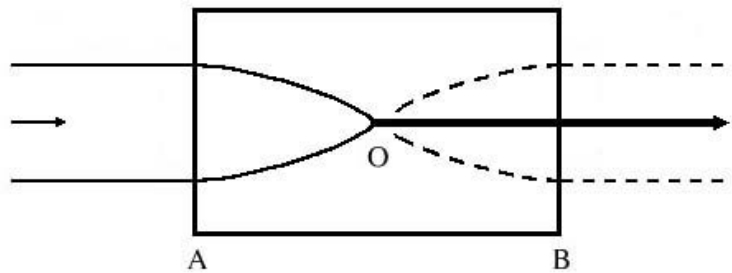

**Fig 6. Self-focusing process of light**

## 4.4 Double and multi-photon absorption[4]

In double absorption process of photons, an electron in low-energy level would absorb two photons at frequencies $\omega_1$ and $\omega_2$ and transits to high- energy level. But if the electron at high-energy level transits back to low-energy level, it either gives out only a photon at frequency $\omega_3 = \omega_1 + \omega_2$, or two photons at frequencies $\omega_1' \neq \omega_1$, $\omega_2' \neq \omega_2$. In general it would not give out two photons with original frequencies $\omega_1$ and $\omega_2$. So double photon absorption processes are irreversible. And so do for multi-photon absorption.

## 4.5 Photon echo phenomena[5]

Under a certain conditions of temperature and magnetic field, a beam of laser can be resolved into two lights with a time difference by using a time regulator of optics. Then two lights are emitted into a proper crystal. Thus three light signs can be observed when they pass through the crystal. The last sign is photon echo. This is a kind of instant coherent phenomena of light. If these three lights signs are imported into same medium again, they can't return into origin two lights. Either three signs are observed (no now echo is caused) or more signs are observed (there are new signs are caused). In fact, besides photon echo, there are electron spin echo, ferromagnetic echo and plasma echo and so on. All of them are irreversible and violate time reversal symmetry.

## 4.6 Light's spontaneous radiation processes

As we known that there exists two kinds of different processes for light's radiations, i.e., spontaneous radiation and stimulated radiation, but there exists only one kind of absorption process, i.e., stimulated radiation without spontaneous absorption in nature. An electron can only transform from high energy level into low energy level by emitting a photon spontaneously, but it can not transform from low energy level into high energy level by absorbing a photon spontaneously. So the processes of light's absorptions themselves are obviously asymmetrical under time reversal.

## 5. Retarded interaction and non-equilibrium correction function in statistical physics

In Document (1), the author had discussed the reason that symmetry violation of time reversal in the high order stimulated radiation and absorption of light. In view of this problem's significance, we discuss this problem further in this paper. As we known that in the current statistical physics, conservative force is used. Based on it, the Liouville equation is symmetrical under time reversal. By considering the fact that the evolution processes of most material systems are irreversible, in order to describe the irreversibility of non-equilibrium systems, we have to introduce some concepts phenomenistically relative to irreversibility and dissipation just as non-equilibrium correlation functions and fluctuation correlation functions and so on. We will point out that the non-equilibrium correlation functions and fluctuation correlation functions are relative to retarded interaction below. For non-equilibrium processes, the introduction of non-equilibrium correction function is essential. As long as it is determined, the non-equilibrium evolution of system would be determined in certain content. In statistical mechanics, the non-equilibrium correlation function is generally defined as [6]

$$G_{AB}(\vec{x},\vec{r},\tau,t) = < A(\vec{x},t)B(\vec{x}+\vec{r},t+\tau) > \tag{24}$$

The formula indicates that the value of physical quantity $A$ at space point $\vec{x}$ and moment $t$ is relative to the value of physical quantity $B$ at space point $\vec{x}+\vec{r}$ and moment $t+\tau$ and the correlative degree is represent by $G_{AB}$. As we known according to special relativity, the propagation speed of interaction is limited. Suppose that the propagation speed of interaction is light's speed in vacuum, let $\tau = -r/c < 0$, the influence of $B$ on $A$ is retarded. If $\tau = r/c > 0$, the influence of $B$ on $A$ is leading. If correlation functions are not local, i.e., for arbitrary point $\vec{x}$, the correlation functions are the same, we can write (24) as

$$G_{AB}(\vec{r},\tau,t) = < A(t)B(\vec{r},t+\tau) > = < A(t)B(\vec{r},t \pm r/c) > \tag{25}$$

Under general situation, (24) and (25) are asymmetrical under time reversal. For equilibrium states, we do the statistical average of (24) over the probability distribution function of equilibrium states, the correlation function can become having nothing to do with time $t$ and coordinate $\vec{x}$ with

$$G_{AB}^0(\vec{r},\tau) = < A(\vec{x},t)B(\vec{x}+\vec{r},t+\tau) >|_0 = < A(\vec{x},t)B(\vec{x}+\vec{r},t \pm r/c) >|_0 \tag{26}$$

Only in this case, the evolution of the system would be symmetry under time reversal. For the quantum statistical processes of non-equilibrium, by considering the non-commutative character of operators, the corresponding non-equilibrium correlation function is defined as

$$G_{AB}(\vec{x},\vec{r},\tau,t) = \frac{1}{2} Tr \hat{\rho} \left[ \hat{A}(\vec{x},t)\hat{B}(\vec{x}+\vec{r},t+\tau) + \hat{B}(\vec{x}+\vec{r},t+\tau)\hat{A}(\vec{x},t) \right] \tag{27}$$

It is the same that the interaction among the different parts is also retarded or leading. Let's discuss a example, i.e., the Casimir-Polder interaction to see the concrete form of correlation function. By considering the fluctuation f uniform electromagnetic field, let $\vec{R} = \vec{r}_2 - \vec{r}_1$, the $i$ partial quantity of dipolar moment at point $\vec{r}_1$ can be written as [7]

$$d_{1i}(\omega) = d_{1i}^{sp} + \alpha_1(\omega)E_i(\vec{r}_1,\omega) + \alpha_1(\omega)D_{ij}^R(\vec{R},\omega)d_{2j}^{sp} + \alpha_1(\omega)D_{ij}^R(\vec{R},\omega)d_{2j}^{sp}\alpha_2(\omega)E_j(\vec{r}_1,\omega) + \cdots \quad (28)$$

Here $D_{ij}^R(\vec{R},\omega)$ is a function relative to the retarded Green's function. The first item on the right side is the spontaneous fluctuation dipolar moment, the second is the dipolar moment induced by fluctuation electric field $E_i(\vec{r}_1,\omega)$, the third is the dipolar moment induced by the retarded electric field which propagates from $\vec{r}_2$ to $\vec{r}_1$, the forth is the dipolar induced by the retarded electric field $E_j(\vec{r}_1,\omega)$ propagating to point $\vec{r}_1$, which is caused by the dipolar moment that is deduced by the fluctuation electric field at point $\vec{r}_2$, and so on. Therefore, the total electric field at point $\vec{r}_1$ is

$$E_i^T(\vec{r},\omega) = E_i(\vec{r}_1,\omega) + D_{ij}^R(\vec{R},\omega)d_{2j}^{sp} + D_{ij}^R(\vec{R},\omega)d_{2j}^{sp}\alpha_2(\omega)E_j(\vec{r}_2,\omega)$$

$$+ D_{ij}^R(\vec{R},\omega)d_{2j}^{sp}\alpha_2(\omega)D_{jk}^R(-\vec{R},\omega)d_{1k}^{sp} + \cdots \quad (29)$$

The first item on the right side of the formula is the fluctuation electric field at $\vec{r}_1$ point. The second is the retarded field that is caused by fluctuation dipolar moment at point $\vec{r}_2$ then is propagated to point $\vec{r}_1$. The third is the retarded electric field that is induced by dipolar moment which is caused by fluctuation electric field at point $\vec{r}_2$ then is propagated to point $\vec{r}_1$, and so on. By using the formulas, we obtain the fluctuation correlation function

$$<\vec{d}_1 \cdot \vec{E}^T(\vec{r},\omega)> = \alpha_1(\omega)\alpha_2(\omega)D_{ij}^R(\vec{R},\omega)<E_i(\vec{r}_1,\omega)E_j(\vec{r}_2,\omega)>$$

$$+ \alpha_1(\omega)D_{ij}^R(\vec{R},\omega)D_{jk}^R(\vec{R},\omega)<d_{2j}^{sp}(\omega)d_{2k}^{sp}(\omega)>$$

$$+ \alpha_1(\omega)\alpha_1(\omega)D_{ij}^R(\vec{R},\omega)<E_i(\vec{r}_1,\omega)_i E_j(\vec{r}_2,\omega)>$$

$$+ \alpha_2(\omega)D_{ij}^R(\vec{R},\omega)D_{jk}^R(-\vec{R},\omega)<d_{1i}^{sp}(\omega)d_{2k}^{sp}(\omega)> + \cdots \quad (30)$$

It is obvious that the correlation function is relative retarded interaction closely. By means of the correlation function, the Casimir-Polder potential can be written as

$$V(R) = -\frac{1}{2}\int_{-\infty}^{\infty}\frac{d\omega}{2\pi}<\vec{d}_1 \cdot \vec{E}^T(\vec{r},\omega)> \quad (31)$$

So the Casimir-Polder potential is also relative to retarded interaction.

We can also use the concept of vacuum fluctuation to deduce the Casimir-Polder interaction energy more simply. Suppose that the vacuum fluctuation electric field of System 2 which located at point $\vec{r}_2$ is composed of $E_0(\vec{r}_2,\omega)$ and $E_1(\vec{r}_2,\omega)$. $E_1(\vec{r}_2,\omega)$ is the retarded electric field induced by the spontaneous dipolar moment $d_1^{sp} = \alpha_1 E_0(\vec{r}_1,\omega)$ at point $\vec{r}_1$ and then is propagated to point $\vec{r}_2$ and $d_1^{sp}$ is caused by the vacuum fluctuation electric field $E_0(\vec{r}_1,\omega)$ at point $\vec{r}_1$. We can let approximately

$$\vec{E}_1(\vec{r}_2,\omega) \approx \frac{\alpha_1 \vec{E}_0(\vec{r}_1,\omega)}{R^3} g\left(\frac{\omega R}{c}\right) \qquad (32)$$

Here $g$ is a dimensionless function. Therefore, we have

$$\vec{E}_2(\vec{r}_2,\omega) = \vec{E}_0(\vec{r}_1,\omega) + \vec{E}_1(\vec{r}_2,\omega) \qquad (33)$$

It is obvious that $\vec{E}_2(\vec{r}_2,\omega)$ is relative to the retarded effect of electric field. The dipolar moment induced by electric field $\vec{E}_2(\vec{r}_2,\omega)$ and interaction energy are individually

$$\vec{d}_2(\vec{r}_2,\omega) = \alpha_0\left[\vec{E}_0(\vec{r}_2,\omega) + \vec{E}_1(\vec{r}_2,\omega)\right] \qquad (34)$$

$$V(\vec{R},\omega) \approx \frac{1}{2}\alpha_2(\omega)\left[\vec{E}_0(\vec{r}_2,\omega) + \vec{E}_1(\vec{r}_2,\omega)\right]^2 \approx \alpha_2(\omega)\vec{E}_0(\vec{r}_2,\omega)\cdot\vec{E}_1(\vec{r}_2,\omega) \qquad (35)$$

Thus, the dipolar moment induced by vacuum fluctuation electric field and the interaction energy of vacuum fluctuation electric field are relative to retarded effects.

By comparing the discussion above and the calculating results of using quantum mechanics and high order retarded effect of radiation field in this paper, we can obtain some conclusion as shown below. In the current non-linear and non-equilibrium statistical mechanics, the irreversibility introduced through fluctuation and non-equilibrium correlation functions are relative closely to retarded interaction. In this way, the calculation can be greatly simplified. However, this kind of method is phenomenistical one, not starts from the original principle, the obtained results would be statistical and approximate ones. There exists some randomicity and uncertainty. For example, only dipolar moments are involved in the formulas above, no multiple moments to exist. All effect of multiple moments is substituted by some parameters which are remained to be determined. But according to the method in this paper, we calculate the problem based on the original principle of quantum physics. Not only dipolar moment, but also multiple moments are involved. We can calculate any high order processes in principle. We can calculate any high order process in principle and all parameters in phenomenistical theory can be determined. So the asymmetry of time reversal deduced in this paper comes from the original dynamic principle of physics with essential significance. The asymmetry violation introduced by non-equilibrium correlation and fluctuation correlation function is caused by statistical method with assumptive element, not coming from original dynamic principle of physics. Though the results deduced from two methods are equivalent, and the method of this paper is complex and difficult comparing with the statistical method.

In fact, the concepts of fluctuation and dissipation are only suitable to the equilibrium or near equilibrium systems. The deduced irreversibility is also limited. For the systems far away from the equilibrium states, it is unsuitable for us using fluctuation theory to explain irreversibility. But the method of this paper is generally suitable. On the other hand, based on the hypothesis of equivalent probability and ensemble, the statistical physics of equilibrium state has been established at present. But as we known that the statistical physics of non-equilibrium state has not been completely established up to new. The reason is that we do not know the dynamic origin of irreversibility in the non-equilibrium processes. All irreversibility in non-equilibrium theory is introduced by the phenomenistical method. Based on the retarded interaction of electromagnetism, we can find the dynamic origin of irreversibility for non-equilibrium statistical physics and reach the unified description for both equilibrium and non-equilibrium physics.

In this paper, we discuss the problems based on non-relative quantum and retarded electromagnetic interaction. The retarded f interaction indicates that the propagation speed of interaction is limited. So after retarded effect is considered, the interaction Hamiltonian is relative one in essence. The only problem is that we have not used quantized operators. By strict consideration, we should discuss problem based on quantum field theory. Because the Hamiltonian of electromagnetic interaction is symmetrical under time reversal, in light of the common understanding at present, the processes of light's stimulated and absorption would be symmetry under time reversal. Then, whether or not the results obtained in the paper that the processes of light's high order stimulated and absorption and non-linear optics violate time reversal symmetry contradicts with the judgment of quantum field theory?

The author has explained this problem in Document (1). The reason is that in the interaction between bounding atoms and photons, the energy levels of bounding atoms are not continuous. By the restriction of energy conservation law, certain filial or partial transition processes are forbidden or can't be achieved. These restrictions would cause the symmetry violation of time reversal of other partial transition processes which can be actualized really. These realizable filial or partial processes which violate time reversal symmetry generally are just the practically observable physical processes. For the electromagnetic interaction between non-bounding atoms and radiation fields, there is no this kind of symmetry violation of time reversal. The calculations show that the violation of time reversal takes place in the second order process. So when the strength of radiation field is big enough, the violation of symmetry would be great. But if all realizable and forbidden processes are taken into account, the total processes of electromagnetic interaction are still symmetrical under time reversal. So the results in this paper do not contradict with the judgment of quantum field theory.

On the other hand, in quantum field theory, w can discuss time reversal in both coordinate space and momentum space. The interaction Hamiltonian in the coordinate space contains all possible processes actually, so the symmetry of time reversal is also for the sum of all processes. In the momentum space, however, we calculate concrete filial or partial transition processes, in stead of total processes actually. It can be proved in light of quantum field theory that the normalization process of the third order vertex angles in momentum space for single process also violates time reversal symmetry, though the sum of all filial processes is still symmetric in the coordinate space. However, the violation of time reversal in the normalization process of the third order vertex angles is very small only with the magnitude of $10^{-5}$, we can not find them in light of the current precision of experiments with the magnitude of $10^{-3}$ at most. We need more accurate experiments to verify the existence of symmetry violation of time reversal in the high order normalization processes of quantum field theory. This problem is neglected at present. The author will discuss this problem in detail later.

The author gratefully acknowledges the valuable discussions with Professors Qiu Yishen in Physical and Optical Technology College, Fujian Normal University and Zheng Shibiao in Physical Department, Fuzhou University.

# Appendix

The transition probability amplitude of the third order process for light's stimulated radiation and absorption

$$a_m^{(3)}(t) = \frac{1}{\hbar^2} \sum_n \left\{ \frac{\hat{F}_{2mn} \hat{F}_{1nl}}{\omega + \omega_{nl}} \left[ \frac{e^{i(3\omega + \omega_{nl} + \omega_{mn})t} - 1}{3\omega + \omega_{nl} + \omega_{mn}} - \frac{e^{i(2\omega + \omega_{mn})t} - 1}{2\omega + \omega_{mn}} \right] \right.$$

$$- \frac{\hat{F}_{2mn} \hat{F}_{1nl}^+}{\omega - \omega_{nl}} \left[ \frac{e^{i(\omega + \omega_{nl} + \omega_{mn})t} - 1}{\omega + \omega_{nl} + \omega_{mn}} - \frac{e^{i(2\omega + \omega_{mn})t} - 1}{2\omega + \omega_{mn}} \right]$$

$$- \frac{\hat{F}_{2mn}^+ \hat{F}_{1nl}}{\omega - \omega_{nl}} \left[ \frac{e^{-i(\omega - \omega_{nl} - \omega_{mn})t} - 1}{\omega - \omega_{nl} - \omega_{mn}} - \frac{e^{-i(2\omega - \omega_{mn})t} - 1}{2\omega - \omega_{mn}} \right]$$

$$+ \frac{\hat{F}_{2mn}^+ \hat{F}_{1nl}^+}{\omega - \omega_{nl}} \left[ \frac{e^{-i(3\omega - \omega_{nl} - \omega_{mn})t} - 1}{3\omega - \omega_{nl} - \omega_{mn}} - \frac{e^{-i(2\omega - \omega_{mn})t} - 1}{2\omega - \omega_{mn}} \right]$$

$$+ \frac{\hat{F}_{0mn} \hat{F}_{1nl}}{\omega + \omega_{nl}} \left[ \frac{e^{i(\omega + \omega_{nl} + \omega_{mn})t} - 1}{\omega + \omega_{nl} + \omega_{mn}} - \frac{e^{i\omega_{mn} t} - 1}{\omega_{mn}} \right]$$

$$+ \frac{\hat{F}_{0mn} \hat{F}_{1nl}^+}{\omega - \omega_{nl}} \left[ \frac{e^{-i(\omega - \omega_{nl} - \omega_{mn})t} - 1}{\omega - \omega_{nl} - \omega_{mn}} + \frac{e^{i\omega_{mn} t} - 1}{\omega_{mn}} \right]$$

$$+\frac{\hat{F}_{1mn}\hat{F}_{2nl}}{2\omega+\omega_{nl}}\left[\frac{e^{i(3\omega+\omega_{nl}+\omega_{mn})t}-1}{3\omega+\omega_{nl}+\omega_{mn}}-\frac{e^{i(\omega+\omega_{mn})t}-1}{\omega+\omega_{mn}}\right]$$

$$+\frac{\hat{F}_{1mn}\hat{F}_{2nl}^{+}}{2\omega-\omega_{nl}}\left[\frac{e^{-i(\omega-\omega_{nl}-\omega_{mn})t}-1}{\omega-\omega_{nl}-\omega_{mn}}+\frac{e^{i(\omega+\omega_{mn})t}-1}{\omega+\omega_{mn}}\right]$$

$$+\frac{\hat{F}_{0nl}\hat{F}_{1mn}}{\omega_{nl}}\left[\frac{e^{i(\omega+\omega_{nl}+\omega_{mn})t}-1}{\omega+\omega_{nl}+\omega_{mn}}-\frac{e^{i(\omega+\omega_{mn})t}-1}{\omega+\omega_{mn}}\right]$$

$$+\frac{\hat{F}_{1mn}^{+}\hat{F}_{2nl}}{2\omega+\omega_{nl}}\left[\frac{e^{i(\omega+\omega_{nl}+\omega_{mn})t}-1}{\omega+\omega_{nl}+\omega_{mn}}+\frac{e^{-i(\omega-\omega_{mn})t}-1}{\omega-\omega_{mn}}\right]$$

$$-\frac{\hat{F}_{1mn}^{+}\hat{F}_{2nl}^{+}}{2\omega-\omega_{nl}}\left[\frac{e^{-i(3\omega-\omega_{nl}-\omega_{mn})t}-1}{3\omega-\omega_{nl}-\omega_{mn}}-\frac{e^{-i(\omega-\omega_{mn})t}-1}{\omega-\omega_{mn}}\right]$$

$$-\frac{\hat{F}_{0nl}\hat{F}_{1mn}^{+}}{\omega_{nl}}\left[\frac{e^{-i(\omega-\omega_{nl}-\omega_{mn})t}-1}{\omega-\omega_{nl}-\omega_{mn}}-\frac{e^{-i(\omega-\omega_{mn})t}-1}{\omega-\omega_{mn}}\right]$$

$$+\frac{1}{\hbar^{3}}\sum_{n,k}\left\{-\frac{\hat{F}_{1mn}\hat{F}_{1nk}\hat{F}_{1kl}}{\omega+\omega_{kl}}\left[\frac{e^{i(3\omega+\omega_{nl}+\omega_{mn})t}-1}{(2\omega+\omega_{kl}+\omega_{nk})(3\omega+\omega_{kl}+\omega_{nk}+\omega_{mn})}\right.\right.$$

$$-\frac{e^{i(\omega+\omega_{mn})t}-1}{(2\omega+\omega_{kl}+\omega_{nk})(\omega+\omega_{mn})}-\frac{e^{i(2\omega+\omega_{nk}+\omega_{mn})t}-1}{(2\omega+\omega_{nk}+\omega_{mn})(\omega+\omega_{nk})}$$

$$+\frac{e^{i(\omega+\omega_{mn})t}-1}{(\omega+\omega_{nk})(\omega+\omega_{mn})}\right]+\frac{\hat{F}_{1mn}\hat{F}_{1nk}\hat{F}_{1kl}^{+}}{\omega-\omega_{kl}}\left[\frac{e^{i(\omega+\omega_{kl}+\omega_{nk}+\omega_{mn})t}-1}{(\omega_{kl}+\omega_{nk})(\omega+\omega_{kl}+\omega_{nk}+\omega_{mn})}\right.$$

$$-\frac{e^{i(\omega+\omega_{mn})t}-1}{(\omega_{kl}+\omega_{nk})(\omega+\omega_{mn})}-\frac{e^{i(2\omega+\omega_{nk}+\omega_{mn})t}-1}{(\omega+\omega_{nk})(2\omega+\omega_{nk}+\omega_{mn})}+\frac{e^{i(\omega+\omega_{mn})t}-1}{(\omega+\omega_{nk})(\omega+\omega_{mn})}\right]$$

$$-\frac{\hat{F}_{1mn}\hat{F}_{1nk}^{+}\hat{F}_{1kl}^{+}}{\omega+\omega_{kl}}\left[\frac{e^{i(\omega+\omega_{kl}+\omega_{nk}+\omega_{mn})t}-1}{(\omega_{kl}+\omega_{mn})(\omega+\omega_{kl}+\omega_{nk}+\omega_{mn})}\right.$$

$$-\frac{e^{i(\omega+\omega_{mn})t}-1}{(\omega_{kl}+\omega_{nk})(\omega+\omega_{mn})}+\frac{e^{i(\omega_{nk}+\omega_{mn})t}-1}{(\omega-\omega_{nk})(\omega_{nk}+\omega_{mn})}-\frac{e^{i(\omega+\omega_{mn})t}-1}{(\omega-\omega_{nk})(\omega+\omega_{mn})}\right]$$

$$+\frac{\hat{F}_{1mn}\hat{F}_{1nk}^{+}\hat{F}_{1kl}^{+}}{\omega-\omega_{kl}}\left[\frac{e^{-i(\omega-\omega_{kl}-\omega_{nk}-\omega_{mn})t}-1}{(2\omega-\omega_{kl}-\omega_{nk})(\omega-\omega_{kl}-\omega_{nk}-\omega_{mn})}\right.$$

$$+ \frac{e^{i(\omega+\omega_{mn})t}-1}{(2\omega-\omega_{kl}-\omega_{nk})(\omega+\omega_{mn})} + \frac{e^{i(\omega_{nk}+\omega_{mn})t}-1}{(\omega-\omega_{nk})(\omega_{nk}+\omega_{mn})} - \frac{e^{i(\omega+\omega_{mn})t}-1}{(\omega-\omega_{nk})(\omega+\omega_{mn})}\Bigg]$$

$$- \frac{\hat{F}_{1mn}^+ \hat{F}_{1nk} \hat{F}_{1kl}}{\omega+\omega_{kl}} \Bigg[ \frac{e^{i(\omega+\omega_{kl}+\omega_{nk}+\omega_{mn})t}-1}{(\omega_{kl}+\omega_{nk})(\omega+\omega_{kl}+\omega_{nk}+\omega_{mn})}$$

$$+ \frac{e^{-i(\omega-\omega_{mn})t}-1}{(2\omega+\omega_{kl}+\omega_{nk})(\omega-\omega_{mn})} - \frac{e^{i(\omega_{nk}+\omega_{mn})t}-1}{(\omega+\omega_{nk})(\omega_{nk}+\omega_{mn})} - \frac{e^{-i(\omega-\omega_{mn})t}-1}{(\omega+\omega_{nk})(\omega-\omega_{mn})}\Bigg]$$

$$- \frac{\hat{F}_{1mn}^+ \hat{F}_{1nk} \hat{F}_{1kl}^+}{\omega-\omega_{kl}} \Bigg[ \frac{e^{i(\omega-\omega_{kl}-\omega_{nk}-\omega_{mn})t}-1}{(\omega_{kl}+\omega_{nk})(\omega-\omega_{kl}-\omega_{nk}-\omega_{mn})}$$

$$- \frac{e^{-i(\omega-\omega_{mn})t}-1}{(\omega_{kl}+\omega_{nk})(\omega-\omega_{mn})} + \frac{e^{i(\omega_{nk}+\omega_{mn})t}-1}{(\omega+\omega_{nk})(\omega_{nk}+\omega_{mn})} + \frac{e^{-i(\omega-\omega_{mn})t}-1}{(\omega+\omega_{nk})(\omega-\omega_{mn})}\Bigg]$$

$$+ \frac{\hat{F}_{1mn}^+ \hat{F}_{1nk}^+ \hat{F}_{1kl}}{\omega+\omega_{kl}} \Bigg[ \frac{e^{-i(\omega-\omega_{kl}-\omega_{nk}-\omega_{mn})t}-1}{(\omega_{kl}+\omega_{nk})(\omega-\omega_{kl}-\omega_{nk}-\omega_{mn})}$$

$$- \frac{e^{-i(\omega-\omega_{mn})t}-1}{(\omega_{kl}+\omega_{nk})(\omega-\omega_{mn})} + \frac{e^{-i(2\omega-\omega_{nk}-\omega_{mn})t}-1}{(\omega-\omega_{nk})(2\omega-\omega_{nk}-\omega_{mn})} - \frac{e^{-i(\omega-\omega_{mn})t}-1}{(\omega-\omega_{nk})(\omega-\omega_{mn})}\Bigg]$$

$$+ \frac{\hat{F}_{1mn}^+ \hat{F}_{1nk}^+ \hat{F}_{1kl}^+}{\omega-\omega_{kl}} \Bigg[ \frac{e^{-i(3\omega-\omega_{kl}-\omega_{nk}-\omega_{mn})t}-1}{(2\omega-\omega_{kl}-\omega_{nk})(3\omega-\omega_{kl}-\omega_{nk}-\omega_{mn})}$$

$$- \frac{e^{-i(\omega-\omega_{mn})t}-1}{(2\omega-\omega_{kl}-\omega_{nk})(\omega-\omega_{mn})} - \frac{e^{-i(2\omega-\omega_{nk}-\omega_{mn})t}-1}{(2\omega-\omega_{nk}-\omega_{mn})(\omega-\omega_{nk})} + \frac{e^{-i(\omega-\omega_{mn})t}-1}{(\omega-\omega_{nk})(\omega-\omega_{mn})}\Bigg]$$

The time reversal of transition probability amplitude of third order process of light's stimulated radiation and absorption

$$a_{Tm}^{(3)}(t) = \frac{1}{\hbar^2} \sum_n \Bigg\{ \frac{\hat{F}'_{2nl} \hat{F}'_{1mn}}{\omega+\omega_{mn}} \Bigg[ \frac{e^{i(3\omega+\omega_{nl}+\omega_{mn})t}-1}{3\omega+\omega_{nl}+\omega_{mn}} - \frac{e^{i(2\omega+\omega_{nl})t}-1}{2\omega+\omega_{nl}} \Bigg]$$

$$- \frac{\hat{F}'_{2nl} \hat{F}'^+_{1mn}}{\omega-\omega_{mn}} \Bigg[ \frac{e^{i(\omega+\omega_{nl}+\omega_{mn})t}-1}{\omega+\omega_{nl}+\omega_{mn}} - \frac{e^{i(2\omega+\omega_{nl})t}-1}{2\omega+\omega_{nl}} \Bigg]$$

$$- \frac{\hat{F}'^+_{2nl} \hat{F}'_{1mn}}{\omega+\omega_{mn}} \Bigg[ \frac{e^{-i(\omega-\omega_{nl}-\omega_{mn})t}-1}{\omega-\omega_{nl}-\omega_{mn}} - \frac{e^{-i(2\omega-\omega_{nl})t}-1}{2\omega-\omega_{nl}} \Bigg]$$

$$+\frac{\hat{F}'^{+}_{2nl}\hat{F}'^{+}_{1mn}}{\omega-\omega_{mn}}\left[\frac{e^{-i(3\omega-\omega_{nl}-\omega_{mn})\,t}-1}{3\omega-\omega_{nl}-\omega_{mn}}-\frac{e^{-i(2\omega-\omega_{nl})\,t}-1}{2\omega-\omega_{nl}}\right]$$

$$+\frac{\hat{F}'_{0nl}\hat{F}'_{1mn}}{\omega+\omega_{mn}}\left[\frac{e^{i(\omega+\omega_{nl}+\omega_{mn})\,t}-1}{\omega+\omega_{nl}+\omega_{mn}}-\frac{e^{i\omega_{nl}\,t}-1}{\omega_{nl}}\right]$$

$$-\frac{\hat{F}'_{0nl}\hat{F}'^{+}_{1mn}}{\omega-\omega_{mn}}\left[\frac{e^{-i(\omega-\omega_{nl}-\omega_{mn})\,t}-1}{\omega-\omega_{nl}-\omega_{mn}}-\frac{e^{i\omega_{nl}\,t}-1}{\omega_{nl}}\right]$$

$$+\frac{\hat{F}'_{1nl}\hat{F}'_{2mn}}{2\omega+\omega_{mn}}\left[\frac{e^{i(3\omega+\omega_{nl}+\omega_{mn})\,t}-1}{3\omega+\omega_{nl}+\omega_{mn}}-\frac{e^{i(\omega+\omega_{nl})\,t}-1}{\omega+\omega_{nl}}\right]$$

$$+\frac{\hat{F}'_{1nl}\hat{F}'^{+}_{2mn}}{2\omega-\omega_{mn}}\left[\frac{e^{-i(\omega-\omega_{nl}-\omega_{mn})\,t}-1}{\omega-\omega_{nl}-\omega_{mn}}+\frac{e^{i(\omega+\omega_{nl})\,t}-1}{\omega+\omega_{nl}}\right]$$

$$+\frac{\hat{F}'_{0mn}\hat{F}'_{1nl}}{\omega_{mn}}\left[\frac{e^{i(\omega+\omega_{nl}+\omega_{mn})\,t}-1}{\omega+\omega_{nl}+\omega_{mn}}-\frac{e^{i(\omega+\omega_{nl})\,t}-1}{\omega+\omega_{nl}}\right]$$

$$+\frac{\hat{F}'^{+}_{1nl}\hat{F}'_{2mn}}{2\omega+\omega_{mn}}\left[\frac{e^{i(\omega+\omega_{nl}+\omega_{mn})\,t}-1}{\omega+\omega_{nl}+\omega_{mn}}+\frac{e^{-i(\omega-\omega_{nl})\,t}-1}{\omega-\omega_{nl}}\right]$$

$$+\frac{\hat{F}'^{+}_{1nl}\hat{F}'^{+}_{2mn}}{2\omega-\omega_{mn}}\left[\frac{e^{-i(3\omega-\omega_{nl}-\omega_{mn})\,t}-1}{3\omega-\omega_{nl}-\omega_{mn}}-\frac{e^{-i(\omega-\omega_{nl})\,t}-1}{\omega-\omega_{nl}}\right]$$

$$-\frac{\hat{F}'_{0mn}\hat{F}'^{+}_{1nl}}{\omega_{mn}}\left[\frac{e^{-i(\omega-\omega_{nl}-\omega_{mn})\,t}-1}{\omega-\omega_{nl}-\omega_{mn}}-\frac{e^{-i(\omega-\omega_{nl})\,t}-1}{\omega-\omega_{nl}}\right]\bigg\}$$

$$+\frac{1}{\hbar^{3}}\sum_{n,k}\Bigg\{-\frac{\hat{F}'_{1nl}\hat{F}'_{1mk}\hat{F}'_{1kn}}{\omega+\omega_{mk}}\left[\frac{e^{i(3\omega+\omega_{mk}+\omega_{kn}+\omega_{nl})\,t}-1}{(2\omega+\omega_{mk}+\omega_{kn})(3\omega+\omega_{mk}+\omega_{kn}+\omega_{nl})}\right.$$

$$-\frac{e^{i(\omega+\omega_{nl})\,t}-1}{(2\omega+\omega_{mk}+\omega_{kn})(\omega+\omega_{nl})}-\frac{e^{i(2\omega+\omega_{kn}+\omega_{nl})\,t}-1}{(\omega+\omega_{kn})(2\omega+\omega_{kn}+\omega_{nl})}$$

$$+\frac{e^{i(\omega+\omega_{nl})\,t}-1}{(\omega+\omega_{kn})(\omega+\omega_{nl})}\Bigg]-\frac{\left(\hat{F}'^{+}_{1nl}\right)^{*}\hat{F}'^{+}_{1mk}\left(\hat{F}'^{+}_{1kn}\right)^{*}}{\omega-\omega_{mk}}\left[\frac{e^{i(\omega+\omega_{mk}+\omega_{kn}+\omega_{nl})\,t}-1}{(\omega_{mk}+\omega_{kn})(\omega+\omega_{mk}+\omega_{kn}+\omega_{nl})}\right.$$

$$-\frac{e^{i(\omega+\omega_{nl})\,t}-1}{(\omega_{mk}+\omega_{kn})(\omega+\omega_{nl})}-\frac{e^{i(2\omega+\omega_{kn}+\omega_{nl})\,t}-1}{(\omega+\omega_{kn})(2\omega+\omega_{kn}+\omega_{nl})}+\frac{e^{i(\omega+\omega_{nl})\,t}-1}{(\omega+\omega_{kn})(\omega+\omega_{nl})}\Bigg]$$

$$
\begin{aligned}
&+ \frac{\hat{F}'_{1nl}\hat{F}'_{1mk}\hat{F}'^{+}_{1kn}}{\omega+\omega_{mk}} \Bigg[ \frac{e^{i(\omega+\omega_{mk}+\omega_{kn}+\omega_{nl})\,t}-1}{(\omega_{mk}+\omega_{kn})(\omega+\omega_{mk}+\omega_{kn}+\omega_{nl})} \\
&- \frac{e^{i(\omega+\omega_{nl})\,t}-1}{(\omega_{mk}+\omega_{kn})(\omega+\omega_{nl})} + \frac{e^{i(\omega_{kn}+\omega_{nl})\,t}-1}{(\omega-\omega_{kn})(\omega_{kn}+\omega_{nl})} - \frac{e^{i(\omega+\omega_{nl})\,t}-1}{(\omega-\omega_{kn})(\omega+\omega_{nl})} \Bigg] \\
&- \frac{\hat{F}'_{1nl}\hat{F}'^{+}_{1mk}\hat{F}'^{+}_{1kn}}{\omega-\omega_{mk}} \Bigg[ \frac{e^{-i(\omega-\omega_{mk}-\omega_{kn}-\omega_{nl})\,t}-1}{(2\omega-\omega_{mk}-\omega_{kn})(\omega-\omega_{mk}-\omega_{kn}-\omega_{nl})} \\
&+ \frac{e^{i(\omega+\omega_{nl})\,t}-1}{(2\omega-\omega_{mk}-\omega_{kn})(\omega+\omega_{nl})} + \frac{e^{i(\omega_{nk}+\omega_{nl})\,t}-1}{(\omega-\omega_{kn})(\omega_{kn}+\omega_{nl})} - \frac{e^{i(\omega+\omega_{nl})\,t}-1}{(\omega-\omega_{kn})(\omega+\omega_{nl})} \Bigg] \\
&- \frac{\hat{F}'^{+}_{1nl}\left(\hat{F}'^{+}_{1mk}\right)^{*}\hat{F}'^{+}_{1kn}}{\omega+\omega_{mk}} \Bigg[ \frac{e^{-i(\omega-\omega_{mk}-\omega_{kn}-\omega_{nl})\,t}-1}{(\omega_{mk}+\omega_{kn})(\omega-\omega_{mk}-\omega_{kn}-\omega_{nl})} \\
&- \frac{e^{-i(\omega-\omega_{nl})\,t}-1}{(\omega_{mk}+\omega_{kn})(\omega-\omega_{nl})} + \frac{e^{-i(2\omega-\omega_{kn}-\omega_{nl})\,t}-1}{(\omega-\omega_{kn})(2\omega-\omega_{kn}-\omega_{nl})} - \frac{e^{-i(\omega-\omega_{nl})\,t}-1}{(\omega-\omega_{kn})(\omega-\omega_{nl})} \Bigg] \\
&+ \frac{\hat{F}'^{+}_{1nl}\left(\hat{F}'^{+}_{1mk}\right)^{*}\left(\hat{F}'^{+}_{1kn}\right)^{*}}{\omega+\omega_{mk}} \Bigg[ \frac{e^{i(\omega+\omega_{mk}+\omega_{kn}+\omega_{nl})\,t}-1}{(2\omega+\omega_{mk}+\omega_{kn})(\omega+\omega_{mk}+\omega_{kn}+\omega_{nl})} \\
&+ \frac{e^{-i(\omega-\omega_{nl})\,t}-1}{(2\omega+\omega_{mk}+\omega_{kn})(\omega-\omega_{nl})} - \frac{e^{i(\omega_{kn}+\omega_{nl})\,t}-1}{(\omega+\omega_{kn})(\omega_{kn}+\omega_{nl})} - \frac{e^{-i(\omega-\omega_{nl})\,t}-1}{(\omega+\omega_{kn})(\omega-\omega_{nl})} \Bigg] \\
&+ \frac{\hat{F}'^{+}_{1nl}\hat{F}'^{+}_{1mk}\hat{F}'_{1kn}}{\omega-\omega_{mk}} \Bigg[ \frac{e^{-i(\omega-\omega_{mk}-\omega_{kn}-\omega_{nl})\,t}-1}{(\omega_{mk}+\omega_{kn})(\omega-\omega_{mk}-\omega_{kn}-\omega_{nl})} \\
&- \frac{e^{-i(\omega-\omega_{nl})\,t}-1}{(\omega_{mk}+\omega_{kn})(\omega-\omega_{nl})} + \frac{e^{i(\omega_{kn}+\omega_{nl})\,t}-1}{(\omega+\omega_{kn})(\omega_{kn}+\omega_{nl})} + \frac{e^{-i(\omega-\omega_{nl})\,t}-1}{(\omega+\omega_{kn})(\omega-\omega_{nl})} \Bigg] \\
&- \frac{\hat{F}'^{+}_{1nl}\hat{F}'^{+}_{1mk}\hat{F}'^{+}_{1kn}}{\omega-\omega_{mk}} \Bigg[ \frac{e^{-i(3\omega-\omega_{mk}-\omega_{kn}-\omega_{nl})\,t}-1}{(2\omega-\omega_{mk}-\omega_{kn})(3\omega-\omega_{mk}-\omega_{kn}-\omega_{nl})} \\
&- \frac{e^{-i(\omega-\omega_{nl})\,t}-1}{(2\omega-\omega_{mk}-\omega_{kn})(\omega-\omega_{nl})} - \frac{e^{-i(2\omega-\omega_{kn}-\omega_{nl})\,t}-1}{(2\omega-\omega_{kn}-\omega_{nl})(\omega-\omega_{kn})} + \frac{e^{-i(\omega-\omega_{nl})\,t}-1}{(\omega-\omega_{kn})(\omega-\omega_{nl})} \Bigg] \Bigg\}
\end{aligned}
$$